# "Predicting" after peeking into the future: Correcting a fundamental flaw in the SAOM - TERGM comparison of Leifeld & Cranmer (2019)[*]

Authors: Per Block[a], James Hollway[b], Christoph Stadtfeld[a], Johan Koskinen[c] & Tom Snijders[d]

[a.] ETH Zürich, Switzerland
[b.] Graduate Institute Geneva, Switzerland
[c.] University of Melbourne, Australia
[d.] University of Groningen, Netherlands

**Abstract**: We review the empirical comparison of SAOMs and TERGMs by Leifeld & Cranmer (2019) in *Network Science*. We note that their model specification uses nodal covariates calculated from observed degrees instead of using structural effects, thus turning endogeneity into circularity. In consequence, their out-of-sample predictions using TERGMs are based on out-of-sample information and thereby predict the future using observations from the future. We conclude that their analysis rest on erroneous model specifications that render the article's conclusions meaningless. Consequently, researchers should disregard recommendations from the criticized paper when making informed modelling choices.

## 1. Introduction

In their recent article '*A theoretical and empirical comparison of the temporal exponential random graph model and the stochastic actor-oriented model*', Leifeld and Cranmer (L&C) compare two statistical methods for the analysis of longitudinal network data: the Stochastic Actor-Oriented Model (SAOM) and the Temporal Exponential Random Graph Model (TERGM)[1]. They present a theoretical discussion and a simulation study as well as an application to an empirical data set. While many parts of their theoretical discussion are contestable (see Section 5 of this response), we focus here on the empirical application. In this empirical study, L&C claim that "when considering out-of-sample predictive performance, the TERGM outperformed the SAOM by a substantial margin," and concluded "that the need of the SAOM to have its updating assumptions met with a high degree of precision is essential in order for that specific model to outperform the more general TERGM" (p. 46). They suggests that, in many cases, the TERGM should be the method of choice.

---

[*] We thank Christian Steglich, Viviana Amati and Felix Schönenberger for feedback on various fronts. Closely related issues about statistical network modelling were brought up at the annual Duisterbelt meetings; we thank all participants that contributed to these discussions. The authors declare no competing interests.

[1] Note that L&C's TERGM is unrelated to the many other discrete-time ERGM models proposed in the literature such as e.g. Krivitsky and Handcock (2014) that is implemented in the 'tergm' R package (Krivitsky and Handcock, 2017).

Upon reviewing the replication material for which the authors kindly sent us the coordinates (Leifeld and Cranmer, 2019b), we found that in their predictive out-of-sample analysis, L&C extract and use substantial information from the test data, i.e. the observation of the network in the future, when generating their ("out-of-sample") predictions – despite multiple claims to the contrary (p.35, p.43). Thus, their predictions used for model evaluation are not "out-of-sample", but built on central, observed features of the network they claim to predict. Thus, what the paper actually shows is that by using partial knowledge of the future, their TERGM variant can adequately recover closely related features of the future. The comparison case using the SAOM makes no such use of future information when simulating beyond the observed data, rendering the comparison meaningless.

The specific error L&C make in their TERGM specification is that, using a network data set that is a four-wave panel, wishing to predict the fourth wave (the 'test set') based on a model estimated on the first three waves (the 'training set'), they add the vectors of indegrees and outdegrees for the fourth wave as exogenous nodal covariates to the training set. Evidently, adding part of the test set to the training set is a circularity in modeling that destroys the endogeneity of the model – and any claims to it being generative. When rectifying these misconceptions and replacing the questionable terms with classically used ERGM statistics that model the same tendencies as endogenous, their claims that the TERGM "outperforms" another model falls apart – and with it the contribution of the empirical part of their study.

In this article we first provide a summary and explanations of our findings for an audience with basic knowledge about statistical network models in Section 2. In Section 3 & 4, we show the technical details of the analyses and the relevant code used by L&C to support our presented conclusions. Finally, we point interested readers to further material to gain a better understanding of ERGMs, SAOMs and related models.

## 2. Summary

### 2.1. Unpacking the empirical analysis by L&C: the circularity

The longitudinal method employed by L&C has a classical Exponential Random Graph Model (ERGM) at its core. An ERGM is a probability model for networks representing the dependence structure between the network ties by endogenous terms and exogenous covariates. The TERGM is a model for network panel data where the $t$'th network observation is modeled by an ERGM in which functions of the preceding network observations can enter as exogenous variables. Various ways to specify this have been proposed (Robins and Pattison, 2001, Hanneke & Xing, 2010; Desmarais & Cranmer, 2012; Krivitsky & Handcock, 2014). In Section 3, we give more information about both models.

The main problem with the analysis by L&C is that it deviates from conventional (T)ERG modelling practices, that even L&C use elsewhere, and this strongly determines their results and makes the comparison meaningless. Prior to their TERGM analysis, the indegree and outdegree of all nodes are calculated at every wave. These sequences of indegrees and outdegrees, after a square root transformation, are then used as *exogenous* nodal covariates in the TERGM analysis for the same wave. The indegrees are used to model the incoming ties of the actors, the outdegrees are used to model outgoing as well as incoming ties, all for the observations for which they were computed. This specification turns endogeneity into circularity, because observed data are used in the model as exogenous nodal covariates, predicting themselves as outcomes.

In comparison to standard ERG modelling, the resulting parameter estimates do not say anything about self-organizing tendencies of the network; neither about dependence between ties, nor about degree centralization of the network. The effect of observed outdegrees on outgoing ties and of observed indegrees on incoming ties is tautological and goes counter to all modeling principles, and should not be used in a reasonable analysis; if used, they must logically have positive parameter estimates. Furthermore, the inclusion of these statistics strongly biases other parameter estimates in the model, due to strong dependencies between the modelled statistics. Thus, that L&C's SAOM and TERGM analysis yield vastly different parameters is not really 'alarming' (p. 42); neither is it true that the "divergence of substantive results [between the SAOM and TERGM] likely has much to do with the fit of these models" (p.42). It is an obvious consequence of the circularity in the model specification. Once we specify a model in which degrees are modelled endogenously (as recommended in introductions to ERG modelling) these differences in parameter estimates disappear, as we demonstrate below.

Thus, the SAOM and the TERGM estimated by L&C do not have "the same specification" (p. 42). Where the SAOM specification has the endogenous indegree popularity (sqrt) effect, their TERGM specification has a circular effect in which observed indegrees (square root transformed) are treated as exogenous covariates; and similarly for two further degree-related effects. The proper analogues of the endogenous degree-related effects in SAOMs for the TERGM analysis would be the endogenous two-star, two-path, and geometrically weighted in-star and out-star effects, which we use in our empirical correction below.

## 2.2. When the future predicts itself

In out-of-sample analysis, a dataset is split into training data and test data. L&C choose the first three waves of the *Knecht network data* (Knecht 2008) as training data and the test data is the fourth wave. Ordinarily, out-of-sample analysis estimates a model using the training data and then data beyond the training data is simulated based on those estimates and compared to the test data. Importantly, the model should use no information from the test data to generate the predictions. However, L&C's "out-of-sample" prediction using the TERGM violates this principle.

We saw above that in the estimation of parameters for each wave the empirically observed degree sequences (for the same wave) were extracted and used as exogenous nodal covariates. The same was done to generate likely outcomes of wave 4 (out-of-sample prediction). The observed indegrees and outdegrees of wave 4 were used, after a square root transformation, as nodal covariates to predict the network in wave 4. The circularity here is evident, perhaps even more so than in the preceding section about modeling, because now part of the test data is used to predict the test data. Clearly, this approach cannot be regarded as out-of-sample testing in the sense outlined in the previous paragraph. The comparison with the SAOM results with respect to prediction of wave 4 ties is meaningless, because the SAOM predictions do not use information from wave 4.

It would in principle be possible that a properly specified TERGM could perform very well in an out-of-sample analysis, supporting L&C's claims. To test this, we estimated a TERGM specified in line with the principles of and literature on ERG modelling, that is, with endogenous degree parameters, to generate out-of-sample predictions. While details are outlined in Section 4, we find that the performance is no better than that of an analogous SAOM[2], corroborating previous analyses that came to the same conclusion (Block *et al.* 2018). Thus, L&Cs claim rests fully on using future information in their analysis. We believe we have shown sufficiently that the conclusions about the questionable idea of one model "outperforming" another by L&C have no grounding, which seems to be their main claim. In the next two sections, we make our arguments more formal and explicit, using L&C's replication material.

---
[2] However, as outlined elsewhere (Block *et al.* 2018) such comparisons have little relevance in the first place, as the best predictive models tend to be primitive and with little explanatory relevance.

## 3. Technical details 1: empirical model specification

In this section, we first outline the differences between ERGMs, TERGMs, and L&C's TERGM-variant more formally. While the TERGM is applicable to multiple waves, we discuss it for two-wave data; extension to further waves is direct. First, we show problems that arise from the model specification used by L&C. Second, we show where in the replication code these errors happen and, third, we show that analysis in line with accepted ERG modelling conventions does not support L&C's paper's conclusions.

### 3.1. Formulation of the model proposed by L&C

We start by considering the ERGM (Wasserman & Pattison, 1996; Robins et al., 2007; Lusher et al., 2013). It is an exponential family model and as such comes with well-known properties. The ERGM defines the probability to observe a network based on statistics of the network. The most commonly used statistics are counts of sub-structures, e.g., the number of reciprocated ties or the number of in-stars of some order, which may be combined with nodal or dyadic attributes. The statistics that are typically used are based on principled assumptions about the dependencies among the ties. The expected prevalence of these statistics is determined by a statistical parameter. The probability to observe the realization *x* of a network is given by

$$p_{\text{ERGM}}(X = x) = \kappa^{-1} \exp\left(\sum_k \theta_k s_k(x)\right),$$

where *X* is the random network state, $\theta$ is a statistical parameter, and *s(x)* a vector of statistics describing the network; $\kappa$ is a normalizing constant. It is important to note for later discussions that the ERGM gives a probability distribution on the set of all networks with the given node set. In empirical cases when we analyze one particular observed network, we aim to find those parameters that make the model probability for the observation as large as possible; this is the maximum likelihood estimate.

The TERGM for two waves at times *t-1* and *t* is defined by the conditional probability function

$$p_{\text{TERGM}}(X(t) = x(t) | X(t-1) = x(t-1)) = \kappa^{-1} \exp\left(\sum_k \theta_k s_k(x(t)) + \sum_h \theta_h z_h(x(t), x(t-1))\right),$$

where $x(t)$ and $x(t-1)$ are the realizations, $X(t)$ and $X(t-1)$ the random networks, $s(x(t))$ a vector of statistics for network $x(t)$ like above, and $z(x(t), x(t-1))$ is a vector of statistics of both networks. For the parameter $\theta$, we denote by $\theta_k$ those pertaining to $s(x(t))$ and by $\theta_h$ those pertaining to $z(x(t), x(t-1))$. The difference with the standard ERGM lie in the extra statistic $z(x(t), x(t-1))$ that are memory terms depending on the networks at time $t$ as well as time $t-1$. A basic example of such a statistic representing the match between the two consecutive observations is the number of identical tie variables,

$$z_h(x(t), x(t-1)) = \sum_{ij} (1 - |x_{ij}(t) - x_{ij}(t-1)|),$$

called the dyadic stability term by L&C. This can be understood as modelling that ties will have some inertia. If this has a positive parameter $\theta_h$, networks in which there is more overlap with the previous observation are more likely.

Yet, while the above is the model L&C claim to use in their article, the model that they actually use is the following:

$$p_{\text{TERGM; LC}}(X(t) = x(t)|X(t-1) = x(t-1); x_{\text{obs}}) =$$

$$= \kappa^{-1} \exp\left(\sum_k \theta_k s_k(x(t)) + \sum_h \theta_h z_h(x(t), x(t-1)) + \sum_l \phi_l u_l(x(t), x_{\text{obs}}(t))\right).$$

Here a further set of statistics, denoted $u_l(x(t), x_{\text{obs}}(t))$, is added to the model that is dependent on a particular realization $x_{\text{obs}}(t)$ of the network. In the analysis, this realization is the observed network at time $t=4$ and there are three statistics depending on the observed indegree and outdegree sequences. This is not made explicit in the article, but is apparent when going through the replication materials, as shown below. An intuitive way of conceiving of what is happening is that a crucial summary – vertex degrees – of the observed dependent network is used to predict this same feature of the dependent network.

The issue may appear subtle, because the model that would be analogous to the SAOM specification in question is

$$p_{\text{TERGM}}(X(t) = x(t)|X(t-1) = x(t-1)) =$$

$$= \kappa^{-1} \exp\left(\sum_k \theta_k s_k(x(t)) + \sum_h \theta_h z_h(x(t), x(t-1)) + \sum_l \phi_l u_l(x(t), x(t))\right),$$

where the statistics $u_l(x(t), x_{\text{obs}}(t))$ are replaced by $u_l(x(t), x(t))$. However, this is not a minor issue at all. In the formulation by L&C, a function of the data is calculated and treated as a deterministic exogenous set of values, influencing the random network $x(t)$. In the correct endogenous formulation, the same mathematical function $u_l(.,.)$ is used; this expresses that "$x(t)$ influences itself", i.e., endogeneity. In this correct specification, $u_l(x(t), x(t))$ is a regular endogenous term that could be subsumed under the $s_k(x(t))$.

As an example, one of these terms is referred to by L&C as indegree of alter (sqrt), and specified as an exogenous covariate for the indegrees. The effect of an exogenous variable $b$ on the indegrees is defined by the statistic $u_l(x, b) = \sum_j x_{+j} \sqrt{b_j}$. L&C use this with $b_j = \sqrt{(x_{\text{obs}})_{+j}}$, the square roots of the observed indegrees, yielding $u_l(x(t), x_{\text{obs}}(t)) = \sum_j x_{+j} \sqrt{(x_{\text{obs}})_{+j}}$. If not the observed values but the random network is filled in we obtain $u_l(x(t), x(t)) = \sum_j x_{+j} \sqrt{x_{+j}}$, an endogenous term reflecting indegree variability.

The specification by L&C introduces circularity in the model. What do these unusual additional terms do to their model, as executed, and what interpretation can we draw from it? It is useful to highlight once again that the ERGM gives a probability to observe *any* possible network realization, based on the statistics describing it; and a positive associated parameter means that networks higher on the corresponding statistic are more likely. In consequence, a positive parameter associated to these statistics $u_l(x(t), x_{\text{obs}}(t))$, expressing an aspect of the match between $x(t)$ and $x_{\text{obs}}(t)$, implies that any realization of the network that is closer to the actual observation according to this match has a higher probability to be observed. The term $u_l(x(t), x(t))$, on the contrary, is a structural term representing endogeneity, i.e., the dependence between the network ties in $X(t)$. We continue the example of the indegree of alter (sqrt) term. In the circular specification, this term has the tendency to force the vector of indegrees of the generated networks in a given wave more similar to the observed indegrees (more precisely, their square roots) in the same wave. Of course this will get a positive parameter estimate, because the observed indegrees are indeed (tautologically) similar to themselves. In the endogenous formulation – the in-two-stars, or more appropriately the geometrically weighted indegree or alternating in-stars – this term, when having a positive parameter, has the tendency to lead to a larger variance of the indegrees and with a negative parameter to the opposite.

What do the estimated parameters $\phi_l$ indicate in the estimated models? When including statistics $u_l(x(t), x_{\text{obs}}(t))$ that depend on the *observed* network in the temporal sequence, we only get the answer that networks are likely if they look like the observation. This gives no indication about the salient features or statistical regularities of a network, but only artificially improves the fit of the model to this particular data set without uncovering any dependencies. Taken further, in this spirit we could include the observed network as a dyadic covariate among the predictors – this would lead to perfect explanation in the model, but would have no explanatory value whatsoever (even though this model would not be estimable because of what is called 'separation' in GLM [Albert and Anderson 1984; Santner and Duffy 1986]). Finally, as the degree sequence is used in the estimation of parameters, it also impacts the parameter estimates of other statistics directly through the strong correlations between statistics in network models. Thus, not only are the parameters $\phi_l$ associated with statistics $u_l(x(t), x_{\text{obs}}(t))$ tautological, but also all other model parameters $\theta_k$ and $\theta_h$ will be systematically distorted. In the analysis by L&C, this can be seen in the suspicious finding that friendship dynamics is not transitive.

It is perhaps worth pointing out that the strategy employed by L&C is not the same as conditioning on a degree distribution as, discussed in Snijders and van Duijn (2002).

### 3.2. Code analysis

Here we outline the implementation of the analysis that corresponds to the explanations above. The analyses L&C undertake use the R package btergm. The package 'btergm' builds upon the 'ergm' package in R (not the 'tergm' package). The core functionalities, such as estimation and sampling, are all done within the 'ergm' package. The 'btergm' package adds some wrapper functions that facilitate the possibility to specify the TERGM.

Turning to the analysis in the article, the replication material kindly provided by the authors (Leifeld & Cranmer, 2019b, script `empirical.R`) shows that L&C first create the square root of the indegree and outdegree as attributes of the observed networks (lines 222-230):

```
for (i in 1:length(friendship)) {
  s <- adjust(sex, friendship[[i]])
  friendship[[i]] <- network(friendship[[i]])
  friendship[[i]] <- set.vertex.attribute(friendship[[i]], "sex", s)
  idegsqrt <- sqrt(degree(friendship[[i]], cmode = "indegree"))
  friendship[[i]] <- set.vertex.attribute(friendship[[i]], "idegsqrt", idegsqrt)
  odegsqrt <- sqrt(degree(friendship[[i]], cmode = "outdegree"))
  friendship[[i]] <- set.vertex.attribute(friendship[[i]], "odegsqrt", odegsqrt)
}
```

The R object `friendship` is a list of networks of length 4. The square roots of indegree and outdegree are each stored as an exogenous `vertex.attribute` with the names `"idegsqrt"` and `"odegsqrt"`, respectively. In the estimation of the TERGM for waves 1-3 in lines 243-247 these attributes are now used to model the tendency of nodes to send or receive more ties:

```
tergm.0.firstthree <- mtergm(friendship[1:3] ~ edges + mutual + ttriple
    + transitiveties + ctriple + nodeicov("idegsqrt") + nodeicov("odegsqrt")
    + nodeocov("odegsqrt") + nodeofactor("sex") + nodeifactor("sex")
    + nodematch("sex") + edgecov(primary) + memory("stability"),
    control = control.ergm(MCMC.samplesize = 5000, MCMC.interval = 3000))
```

The three terms `nodeicov("idegsqrt")` + `nodeicov("odegsqrt")` + `nodeocov("odegsqrt")` correspond to the statistics $u_l(x(t), x_{obs}(t))$ we outlined above, and model (i) the tendency of nodes with high *observed* indegree to receive ties, (ii) the tendency of nodes with high *observed* outdegree to receive ties, and (iii) the tendency of nodes with high *observed* outdegree to send ties, respectively. These are exogenous terms, defined for nodal covariates. The first and third of these express, respectively, that indegrees are similar to observed indegrees and that outdegrees are similar to observed outdegrees. These are tautological terms, naturally receiving high parameter estimates. The second term expresses that indegrees are similar to observed outdegrees. This is a circular term at the level of networks, but not a direct tautology, and the resulting parameter estimate is small. It is important to note that the implementation of the `mtergm` function (as well as the `btergm` function) in the 'btergm' package automatically uses these variables at the same time-point as the dependent network, leading to the outlined problems. The term `memory("stability")` models dependence on the previous time-point by the stability term $z_h(x(t), x(t-1))$ described above.

### 3.3. Replication with conventional ERGM terms

In this section, we replicate the analysis by L&C using conventional ERGM terms instead of the circular statistics that try to represent endogenous tendencies by artificially exogenous variables[3]. We do this, partly because the results by L&C suggest that friendships are *not* clustering in groups by transitive closure, contradicting a considerable sum of empirical research, and partly because in our experience, past research that performed both ERGM-type and SAOM-type analyses on the same data tended to yield fairly similar substantive results. In practice, we substitute the terms `nodeicov("idegsqrt")`, `nodeicov("odegsqrt")` and `nodeocov("odegsqrt")` with standard ERGM terms that model degree dispersion, in particular geometrically weighted instars, two-paths and geometrically weighted out-stars (statnet terms `gwidegree`, `twopath` and `gwodegree`). Further, we substitute the triadic terms transitive triplets (`ttriple` and `ctriple`) with their geometrically weighted versions (`dgwesp(type = "OTP")` and `dgwesp(type = "ITP")`). We choose these statistics with their geometrically weighted versions, as the specification of the raw counts is a mis-specified model prone to degeneracy[4].

---

[3] We perform this replication analysis with the code provided by L&C, thus leaving all other modelling and software choices intact.

We further substitute the SAOM terms for transitive and cyclic triplets with the geometrically weighted versions. Despite claims by L&C, the specification in the tutorial article from 2010 are not state-of-the-art ('canonical', p. 42) in 2019, but have evolved considerably. Though model terms in ERGM and SAOM analyses are necessarily different in their dependence assumptions (Block, Stadtfeld and Snijders, 2019), using geometrically weighted versions brings the two analyses close to each other, making the results more comparable.

The results from the analyses are presented in Table 1. In terms of substantive conclusions, the results from both models are remarkably similar. The 'Transitive ties' parameter and the 'GWESP Transitive' parameter need to be interpreted together, as both model the same tendency of transitive closure, but with different functional forms. The combination of parameters shows that both models find a strong tendency towards friendships being transitive. No endogenous sorting of degrees is found, while the impact of the exogenous covariates 'Same primary class' and sex are in the same direction. The only difference is that the 'GWESP cyclic' term is significant in the TERGM analysis but not in the SAOM analysis. This is likely due to the different formulations of the various GWESP and transitivity terms in this ERGM-type and SAOM-type model. We conclude that the substantive insights we can draw from either model are not very different, but that understanding how these differences come about is a more complex task than attributing this to simple differences in model fit. Block *et al.* (2018, 2019) show in similar comparisons how ERGM and SAOM effects differ in their fundamental formulation and that the parameter estimates of a TERGM vary depending on the length of the time period between two subsequent waves even when analyzing time-homogeneous networks evolution.

## 4. Technical details 2: out-of-sample prediction

In the final step of the analysis, L&C perform an out-of-sample test to cross-validate the models. Out-of-sample predictions have been used for validating models for *independent* data in, for example, econometrics. We have outlined elsewhere (Block *et al.*, 2018) why we do not advocate this type of model assessment for network models with highly *interdependent* data though, not least because these

---

[4] Issues with this mis-specified Markov model have been extensively treated in Strauss (1986); Jonasson (1999); Snijders (2002); and Handcock (2003). How these degeneracies are alleviated by different dependence assumptions is covered in Snijders *et al.* (2006) and Schweinberger (2011).

|  | SAOM Analysis | | | TERGM Analysis | | | |
| --- | --- | --- | --- | --- | --- | --- | --- |
|  | est. |  | s.e. | est. |  | s.e. |  |
| Rate period 1 | 8.38 | ° | (1.55) | 0.73 | ° | (0.08) | Memory |
| Rate period 2 | 8.66 | ° | (1.39) |  |  |  |  |
|  |  |  |  |  |  |  |  |
| Outdegree (density) | -1.84 | ° | (0.66) | -2.98 | ° | (0.43) | Edges |
| Reciprocity | 1.61 | *** | (0.28) | 2.03 | *** | (0.37) | Reciprocity |
|  |  |  |  |  |  |  |  |
| Transitive Ties | -0.16 |  | (0.33) | -2.16 | *** | (0.46) | Transitive Ties |
| GWESP Transitive | 1.76 | *** | (0.34) | 2.84 | *** | (0.34) | GWESP Transitive |
| GWESP Cyclic | -0.27 |  | (0.27) | -0.52 | ** | (0.12) | GWESP Cyclic |
|  |  |  |  |  |  |  |  |
| Indegree-popularity (sqrt) | -0.29 |  | (0.27) | 1.07 |  | (0.70) | GW Indegree |
| Outdegree-popularity | -0.10 |  | (0.07) | -0.06 |  | (0.03) | Two-paths |
| Outdegree-activity (sqrt) | 0.01 |  | (0.12) | -0.65 |  | (0.52) | GW Outdegree |
|  |  |  |  |  |  |  |  |
| Same primary class | 0.39 | * | (0.20) | 0.44 | * | (0.17) | Same primary class |
| Boy alter | -0.12 |  | (0.17) | -0.06 |  | (0.15) | Boy alter |
| Boy ego | 0.37 | * | (0.19) | 0.25 | * | (0.13) | Boy ego |
| Same sex | 0.71 | ** | (0.19) | 0.53 | ** | (0.15) | Same sex |

Notes: All analyses performed using standard best practises. Significance levels: * = 0.05; ** = 0.01; *** = 0.001; ° = not tested.

Table 1: Results of TERGM and SAOM analysis using standard endogenous ERGM terms.

types of model generally fare worse than trivial prediction models in out-of-sample predictions. Nevertheless, here we treat the issue on L&C's own terms by investigating how well a conventionally specified TERGM predicts out-of-sample.

In the out-of-sample analysis, the first three waves of the *Knecht data* are used to train the TERGM and the SAOM, as outlined above. Both models allow to simulate future likely outcomes under the model, which allows us to compare expected values of the fourth wave under the model to the actual observation for different fit metrics. One of the metrics used by L&C is tie-prediction. While L&C claim at two different points (p. 35, 43) that their out-of-sample data is generated without using information from the future, their replication code reveals that information from the future is used. The original training model used contemporaneous transformations of the dependent variable (the degree sequence) to explain the dependent variables; carrying this contradiction forward, the predictive model does the same. The in- and out-degree sequences of wave 4 (the test data) are used to generate predictions of the network structure of wave 4. To be very clear: L&C take the network at the fourth wave, compute nodal attributes for this network defined as the square roots of the indegrees and outdegrees, generate samples of the network at the fourth wave based on these attributes, and present this as an out-of-sample prediction. The comparison case, i.e. the out-of-sample predictions of the SAOM, makes no use of any information from the fourth wave, clearly biasing the comparison predictably towards the TERGM, in light of our discussions above.

Based on the biased results, L&C highlight the predictive capabilities of the TERGM, which leads them to suggest that the TERGM might be preferable over the SAOM for analyzing this kind of network data. Because their predicting simulations are contaminated by crucial features of the to-be-predicted data for the TERGM only, their model comparison is meaningless and the conclusions of the article unfounded. In the following sections, we show where in the code in the replication material these errors occur, followed by a more appropriate replication using the model estimated above that relies on classical ERGM terms.

### 4.1. Code analysis

As in the code analysis in Section 3.2, the out-of-sample predictions are generated using the 'btergm' package. One of the wrapper functions of the btergm package is the 'gof' function that allows to simulate out-of-sample predictions for a specified model and, subsequently, calculate goodness-of-fit statistics as introduced by Hunter et al. (2008a). This 'gof' function is prominently applied in section 4 of the paper for the out-of-sample comparison. In particular, the simulation of out-of-sample predictions happens in lines 254-260:

```
tergm.0.oos <- gof(tergm.0.firstthree, nsim = nsim, target = friendship[[4]],
    formula = friendship[3:4] ~ edges + mutual + ttriple + transitiveties
    + ctriple + nodeicov("idegsqrt") + nodeicov("odegsqrt")
    + nodeocov("odegsqrt") + nodeofactor("sex") + nodeifactor("sex")
    + nodematch("sex") + edgecov(primary) + memory("stability"),
    statistics = c(esp, dsp, ideg, geodesic, rocpr), parallel = parallel,
    ncpus = ncpus)
```

The crucial part here is 'formula = friendship[3:4] ~ ...'. This tells the algorithm to use the "covariates" that are stored in the network object friendship[[4]] in the simulations. But as we have seen before, the "covariates" that are used in the model are transformations of the observed network at wave 4. The 'gof' function as used here thus simulates a model using future information (the future in- and out-degree of nodes). This circularity has major consequences for the paper results, as discussed previously. The most sophisticated functions of the ERGM routine (such as sampling and estimation) are not done in the 'btergm' package, but are outsourced to the established 'ergm' package in R (Hunter et al. 2008b). Our code analysis revealed no errors in the 'ergm' package.

The interested reader can verify the use of future information in the simulations without diving into the code of the 'btergm' package the following way. If no information from wave 4 of the `vertex.attributes` with the names `"idegsqrt"` and `"odegsqrt"` would be used in the out-of-sample simulation, manipulating these covariates should make no difference in the distribution of *simulated* networks; in fact, these covariates should change nothing in the analysis whatsoever. However, running the following lines of code before the 'gof' functions dramatically changes the predictions:

```
friendship[[4]] <- set.vertex.attribute(friendship[[4]], "idegsqrt", rep(10,26))
friendship[[4]] <- set.vertex.attribute(friendship[[4]], "odegsqrt", rep(10,26))
```

This code sets the internal covariates of `"idegsqrt"` and `"odegsqrt"` in the fourth wave to a constant value of 10 for all 26 nodes. As the parameter associated to these covariates is strongly positive *and used in the simulations*, setting the attribute values very high results in the simulations generating mostly complete networks. These can be inferred using the 'gof' function in the 'btergm' package. This shows that and how the future is used to generate predictions of the future network state in L&C's analysis.

## 4.2. Replication with classical ERGM terms

Finally, we replicate the out-of-sample cross-validation that was attempted in the original paper using the model specification with classical endogenous ERGM terms that do not use contemporaneous information of the dependent network, as estimated in Section 3.3. Thus, we compare the TERGM and the SAOM without using information from the future in either model, and using specifications of transitivity and cyclicity that are more in line with contemporary usage, but using the same metrics of comparison as L&C did. The results can be found in Figure 1. The GOFs for the auxiliary statistics 'Edge-wise shared partners', 'Dyad-wise shared partners', 'Indegree distribution', and 'Geodesic distances' show no clear trend favoring either model. The ROC and PR curves are, for what they are worth, very similar between the SAOM and the re-specified TERGM. In sum, these analyses indicate no gain in predictive value between a correctly specified TERGM and the SAOM corroborating earlier findings that came to the same conclusion (Block *et al.* 2018).

It would certainly be possible to optimize the model specification of both the TERGM and the SAOM to improve on these metrics, especially as the geometrically weighted terms have an internal parameter that can be adjusted. However, we do not do this here; the aim of this response was to show that there is a deceptive mismatch between the text of the article and the actually undertaken empirical analysis; that these modelling choices that were not laid bare in the text of the article directly influence (indeed explain) the results that were found; that these modelling choices are not in line with contemporary understandings of network dependencies nor with the model specification in the SAOM they were comparing the TERGM to; and that, thus, the conclusions drawn from them are unfounded.

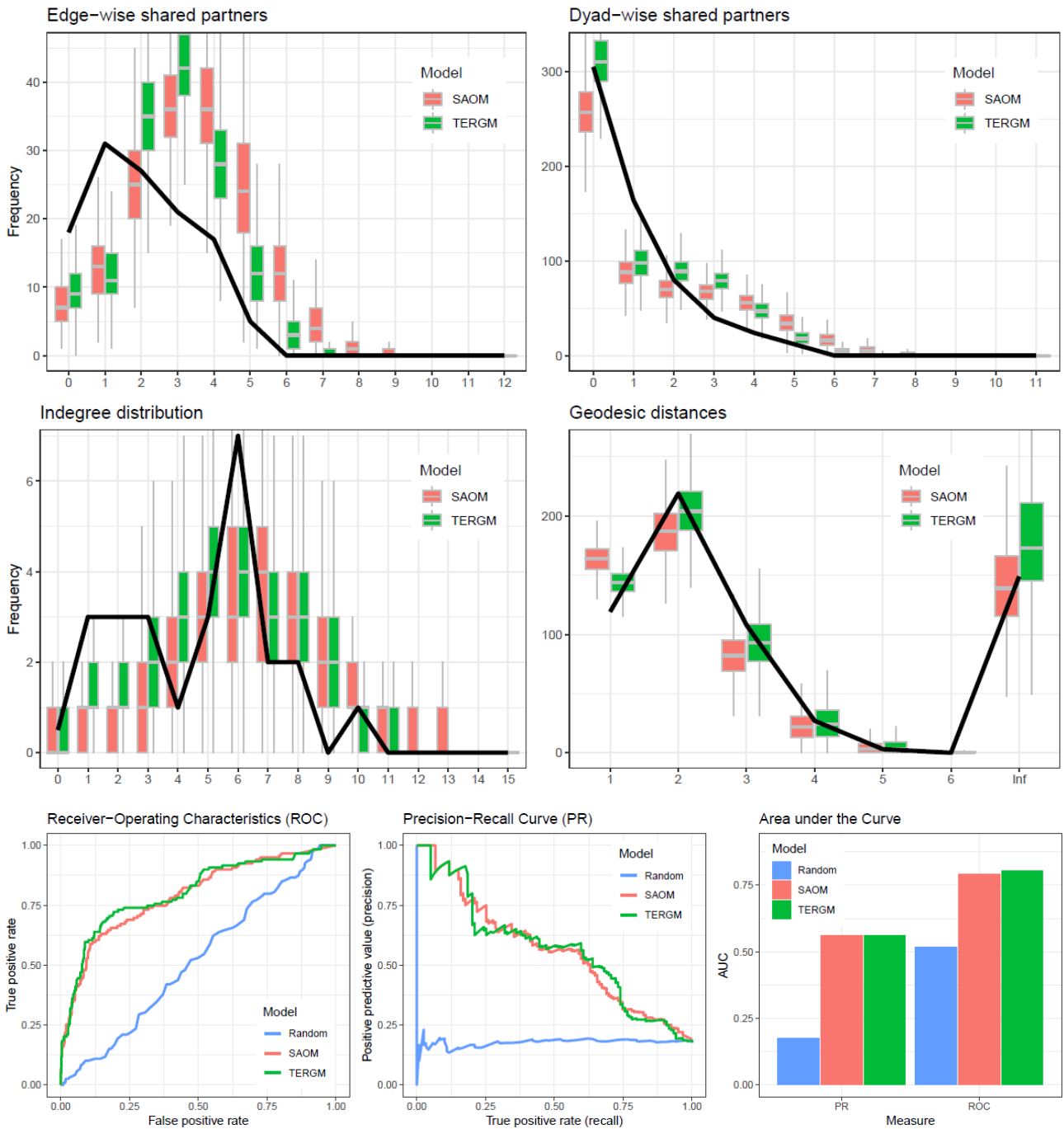

**Figure 1**: Replication of Figure 4 and Figure 5 from L&C's article with ERGM terms in the model that do not use future information. This shows that the claim that TERGMs are superior is false.

# 5. Conclusion and further literature

This response article is a post-publication review of the recent article by Leifeld and Cranmer (2019) in *Networks Science*. While there are some deeper philosophical and theoretical issues with the article, many of these points have already been discussed in other publications: An introduction to the principles of actor-oriented network models are available in Snijders (1996) and Snijders *et al.* (2010). The various aspects of ERG modelling have been treated in Lusher *et al.* (2013), shorter excellent introductions are equally available (e.g. Wasserman & Pattison 1996, Robins *et al.* 2007).

Discussion about estimation methods for SAOMs is given in Snijders (2001, Method of Moments), Koskinen and Snijders (2007, Bayesian), Snijders, Koskinen and Schweinberger (2010, Maximum Likelihood), and Amati *et al.* (2015, Generalized Method of Moments). For ERGMs estimation details for Maximum Likelihood can be found in Snijders (2002) or Geyer and Thompson (1992), and for Pseudo Maximum Likelihood in Strauss and Ikeda (1990). Close reading of these texts, as well as van Duijn, Gile and Handcock (2009) shows the subtle differences between estimation techniques and that the estimates are expected to differ, as well as that that pseudo-maximum likelihood is not a trustworthy method of estimation for the ERGM.

For explicitly comparative articles between models, Schaefer and Marcum (2018) give an introduction to different statistical models for network dynamics where they point out similarities and differences between SAOMs and (S)(T)ERGMs. Going into detail, Block *et al.* (2019) have shown how direct *empirical* comparisons between SAOMs and ERGMs are complicated, as the fundamental dependence assumptions between the models differ in such a way that it is not possible to formulate 'equivalent' models, even if the parameter names might suggest otherwise. In a follow-up, Block *et al.* (2018) focused on the *theoretical* comparison of the TERGM and SAOM considering what L&C call the 'data-generating process' (DGP). They point out that the SAOM is a process-based model and, therefore, its formulation directly proposes a theoretical model how networks evolve (a DGP). The TERGM, in contrast, is a model for a network state; its DGP is a purely technical solution to obtain samples under the model that has no coherent interpretation about a network evolution *process*. Thus, the discussion which model approximates empirical, real-world network evolution better is moot.

This response adds to the articles outlined above in directly engaging with a further comparison by L&C. We show that their empirical comparison between SAOMs and TERGMs on the basis of out-of-sample prediction is distorted: the TERGMs is favoured by using information about the future in their TERGM specification but not in the SAOM specification. Since there are no grounds to use such circular specifications in either model, we compared the original SAOM specification with a more contemporary ERGM specification that uses endogenous effects to model dependencies and found no meaningful difference between the models in terms of prediction. Given that L&C's strong conclusions that the TERGM might generally be preferable are founded on these erroneous model specification, we therefore conclude that the results from L&C's paper can be set aside; this means that applied researchers can return their attention to making principled choices between (statistical) network models for each particular study. We advocate (and demonstrate) such flexibility, as we believe that the various network models to choose from – of which ERGMs and SAOMs are only two examples – can be seen as reflecting the theoretical and substantive diversity in the discipline of social network science.

## 6. Conflicts of Interest

The authors have nothing to disclose.